\documentclass[12pt,aps,prb,preprint]{revtex4}
\usepackage{amsfonts,amsmath,amssymb,graphics,epsfig}
\usepackage{enumerate}
\usepackage{theorem}
\usepackage{paralist}
\numberwithin{equation}{section}

\begin{document}
\title{ {\large {\bf The General Very Special Relativity in Finsler Cosmology}}   }

\author{A.P.Kouretsis }
\affiliation{Department of Physics National Technical University of Athens Zografou Campus  15780  Athens  Greece}
\email{a.kouretsis@yahoo.gr}
 \author{M.Stathakopoulos}
 \affiliation{1 Anastasiou Genadiou str. 11474 Athens Greece}
 \email{michaelshw@yahoo.co.uk}
  \author{P.C.Stavrinos}
 \affiliation{Department of Mathematics University of Athens 15784 Greece}
 \email{pstavrin@math.uoa.gr}

\begin{abstract}
\noindent {\small General Very Special Relativity (GVSR) is the curved space-time of Very Special Relativity (VSR)
proposed by Cohen and Glashow. The geometry of GVSR possesses a line element of Finsler Geometry introduced by Bogoslovsky.
We  calculate the Einstein field equations and derive a modified FRW cosmology for an osculating Riemannian
space. The Friedman equation of motion leads to an explanation of the cosmological acceleration  in terms of an
alternative non-Lorentz invariant theory. A first order approach for a primordial spurionic vector field introduced
into the metric gives back an estimation of the energy evolution and inflation \newline\newline {\bf Key
words}:Finsler Geometry, Cosmology, Lorentz Violations}\newpage
\end{abstract}

\maketitle

\section{Introduction}
It is widely known that relativity violations are arising from breaking the Lorentz symmetry. Lorentz Violations is a very
wide research area traced back to Dirac in early 50s and applied in many aspects of modern Physics since the question of
 Lorentz Symmetry invariance is incorporated to the foundations of both General Relativity and Quantum Field Theory \cite{hist}.
Plenty of different high energy theories face the challenge of Lorentz Symmetry Breaking therefore the proposal of a
different geometrical point of view from General Relativity makes sense. The consideration of such a scenario implies a
class of modified dispersion relations for elementary particles depended on both coordinates and momenta. In this case the
geometry of space-time may be direction-dependent generating a local anisotropy \cite{mavr, girelli, visser}. A possible candidate
for a geometry which incorporates the anisotropies directly to the metric, is Finsler geometry. A typical example of
studying Lorentz violations within the framework of Finsler Geometry is presented in \cite{Bogoslovsky}.

 An interesting case of Lorentz violation where the Finsler Geometry turns up, is the model of Very Special Relativity (VSR)
characterized by a reduced Lorentz symmetry \cite { Coh}. The Lorentz violations are generated by a subgroup of
the full Lorentz group, called $ISIM(2)$. The adoption of this theory is not in contrast to experimental constraints since it appears to be
compatible with all current limits of local Lorentz and CPT invariance, confirming some new physics \cite{Gold, Coh, Dunn}. Some
experimental analysis on the upper bounds of the Lorentz violation are described in \cite{Batt}.

The combination of Lorentz violations to gravitational phenomena is not compatible to the geometrical framework of Riemann
geometry since General Relativity is applicable only to low-energy descriptions of nature. The study of relativity violations
together with gravitational phenomena requires a type of space-time geometry which allows local non-Lorentz invariance while
preserving general coordinate invariance. A direct consequence of Lorentz violations is the production of local anisotropies
and we expect any gravitational phenomenon to be affected by the breaking of classical local flatness (see for example \cite{Kost, Gasp2}). In Finsler Geometry
all geometrical objects are direction-dependent while preserving the general coordinate invariance \cite{RundFB} . Thus
this special type of geometry is a possible choice for the investigation of  geometrodynamics  allowing Lorentz
violations \cite{rwfins,vacaru,vac2,Ish,Asanov} .

 The deformation of the group $ISIM(2)$ leads to the construction of a Finslerian line element proposed by Bogoslovsky
(see \cite{Gibb_fins} and references therein). The whole set of Lorentz transformations are replaced by the deformed
group of transformations $DISIM_b(2)$ which
is a subgroup of the Weyl group. The line element $ds^2=\eta_{ij}dx^i dx^j$ is no-longer preserved under the
$DISIM_b(2)$
transformations . The line element that is  preserved under these transformations is the
non-Riemannian \cite{Bog1}
\begin{equation}
ds= (\eta_{ij}dx^i dx^j)^{(1-b)/2}(n_k dx^k)^b \label{Bog_met}
\end{equation}
The vector field $n^\mu$  is interpreted as a ``spurion vector field'' and it defines the direction of the ``{\ae}theral '' motion's
4-velocity. The dependence of the metric function on the vector $n^\mu$ indicates the anisotropic character of
space-time.
The parameter $b$ is dimensionless and is restricted by various experiments \cite{Bog1,HDlim}. It is inserted
into the mass tensor
\begin{equation}
m_{ij}=(1-b)m(\delta_{ij}+bn_in_j)
\end{equation}
of every particle coming from a Lagrangian constructed in \cite{Gibb_fins}. The canonical momenta satisfies the mass-shell
condition ($\eta_{ij}=\mbox{diag}(+1,-1,-1,-1)$)
\begin{equation}
\eta^{ij}p_i p_j= m^2(1-b^2)\left( \frac{ n^k p_k }{ m(1-b) }  \right)^{2b/(1+b)}.\label{disp}
\end{equation}
which upon quantization leads to the Klein-Gordon equation
 \begin{equation}
 \Box\phi+ m^2(1-b^2) \left(\frac{-in^\mu}{ m(1-b) }\partial_{\mu} \right)^{2b/(1+b)}\phi=0. \label{KGd}
\end{equation}
The dispersion relation (\ref{disp})  implies that the relationship of energy
and momentum is affected by the type of the metric geometry. The {\ae}ther-drift experiments restrict $b$ within the very
tight limits $|b|<10^{-10}$ \:\cite{Bog1} and the anisotropy of
inertia implies $|b|<10^{-26}$ \:\cite{HDlim}.  The Modified Dispersion Relation constructed by Gibbons in \cite{Gibb_fins}
has also been reproduced in a Randers-type Finsler space \cite{Chin_MDR}.

We proceed by constructing the geometrical machinery of space-time using the Finslerian connection and curvature coming
from an osculating Riemann metric \cite{RundFB} . The metric function of the Finsler space is the one prescribed by
Bogoslovsky where the Minkowski metric tensor is substituted by the FRW-metric defined in standard Cosmology. The derivation
of the  gravitational field equations is similar to the one appeared in \cite{rwfins}. The Friedmann equation of motion for
a linearized spurion vector field parallel to the fluid flow lines lead us to a self-accelerated cosmological model. The
kinematical equations of a scalar field are also considered providing an inflationary solution for the scale factor.

\section{The Bogoslovsky's metric applied to Cosmology}
The effects of local Lorentz violation are likely applicable on cosmological contexts, such as those involving the
cosmological constant, dark matter and dark energy. The homogeneous Friedman-Robertson-Walker cosmological solutions may
acquire anisotropic corrections, leading to a realistic anisotropic cosmology complied to the observational data \cite{Kost}.

The Bogoslovsky's metric may shed light on some problems of modern Cosmology which are compatible to local anisotropies of
the geometrical structures of space-time. We can construct a geometrical machinery of cosmology by introducing comoving
coordinates to the metric function
\begin{equation}
F(x,y)= (\eta_{ij}y^i y^j)^{(1-b)/2}(n_k y^k)^b. \label{BogRW_met_fun}
\end{equation}
We replace the Minkowskian  metric with the Robertson-Walker one
\begin{equation}
a_{\mu\nu}=\mbox{diag}\left( 1,-\frac{a^2(t)}{1-kr^2}, -a^2(t)r^2 , -a^2(t)r^2\sin^2\theta \right) \label{RW}
\end{equation}
where $t$ is the cosmic proper time, $r,\theta,\phi$ the comoving spherical coordinates, $k=0,\pm 1$ and $a(t)$ the scale factor of the expanding volume. The new metric function
\begin{equation}
F(x,y)=( a_{\mu\nu} y^\mu y^\nu )^{(1-b)/2 }( n_\rho y^\rho  )^b
\end{equation}
is a direct result of a coordinate linear transformation
\footnote{ $ds^2=  \left( \eta_{ij}dx^idx^j\right)^{b-1}(n_idx^i)^b =\left( \eta_{ij}\frac{\partial x^i}{\partial x^\mu} \frac{\partial x^j}{\partial x^\nu }dx^\mu dx^\nu \right)^{b-1} ( n_i
\frac{ \partial x^i }{ \partial x^\rho } dx^\rho )^b=\left( a_{\mu\nu }dx^\mu dx^\nu\right)^{b-1}( n_\rho dx^\rho)^b $ where the latin and greek indices represent the two coordinate systems}
\begin{equation}
B_{\mu}=\frac{ \partial x^i}{ \partial x^\mu }B_i
\end{equation}
 and it directly determines the metric of the Finslerian space-time
\begin{equation}
f_{\mu\nu}(x,y)=\frac{1}{2}\frac{\partial F^2}{\partial y^\mu \partial y^\nu}(x,y).
\end{equation}
This straightforward generalization for curved space-times and  Machian gravitational theories is used \cite{Bogoslovsky}
to give an explanation of local anisotropies in terms of geometrical phase transitions. The consideration of such
a metric function embodies two types of geometries: the dynamics is described by the Finslerian metric produced by $F(x,y)$
while all the information about gravity is encoded to the FRW metric $a_{\mu\nu}$. A similar formulation is deduced by
Bekenstein in \cite{bekenstein} contemplating a different approach for Finsler Geometry . The variables $y^\mu=\frac{dx^\mu}{dt}$
represent the 4-velocity components of the fluid flow-lines hence
$y^\mu=(1,0,0,0)$.
\newline\newline
{\it A null or time-like spurionic vector field?}\newline
The study of GVSR requires the existence of a null spurionic vector field. However,  this preferential direction of ``$\ae$ther'' is most naturally expected to be tangent
to the flow lines of the cosmological fluid like every primordial vector field \cite{Grav}. Thus, $n^\mu$
must be parallel to the velocity of the comoving observer i.e.
\begin{equation}
n^\mu= \lambda y^\mu .
\end{equation}
Hence, the spurion vector field  becomes  of time-like character  at a late time period of the universe with  $|n^\mu|\ll 1$. Therefore, it is written in coordinate form
\begin{equation}
n^\mu=(n(t),0,0,0)\label{nt}
\end{equation}
with the time component very small. The time-like spurionic vector field does not essentially affect the mass shell condition (\ref{disp}) since only quadratic terms of $n^\mu$ turn up at the contractions of canonical momentum $p_\mu=m\frac{\partial F}{\partial y^\mu}$.
\newline\newline
{\it The osculating space and the gravitational field equations}\newline
All the geometrical quantities of Finsler Geometry depend both on coordinates and velocity.
However, we can study the geometrical properties of a Finsler space by restricting the vector field $y^\mu$ to belong to an
individual tangent space for a given position coordinate. In such a case the velocity coordinates are functions of the
position, therefore we can measure distances by using the  metric \cite{RundFB}
\begin{equation}
g_{\mu\nu}(x)=f_{\mu\nu}(x,y(x)). \label{osc_metr}
\end{equation}
This method, known as the {\it Osculating Riemannian} approach (for details see \cite{RundFB}), can be specialized for  the
tangent vector field $y(x)$ of the cosmological fluid flow lines.

We are interested in producing the Einstein  field equations as in \cite{rwfins}. After calculating the
connection and the curvature for the Riemannian osculating metric $g_{\mu\nu}(x)$ we are lead to \cite{Asanov}
\begin{equation}
L_{\mu\nu}-\frac{1}{2}Lg_{\mu\nu}=-\frac{8\pi G}{c^4}T_{\mu\nu}\label{feqs}
\end{equation}
where all the  quantities in (\ref{feqs}) are functions of  $(x,y),\:y\equiv y(x)$. The energy-momentum tensor for the
signature (+,-,-,-) is defined to be
\begin{equation}
T_{\mu\nu}=-Pg_{\mu\nu}+(\mu+P)y_{\mu}y_{\nu} \label{energy_mom_ten}
\end{equation}
where $P$ is the pressure and $\mu$ is the energy density of an ideal cosmic fluid.

The dispersion relation (\ref{disp}) is also modified after plugging the FRW metric and the vector $n^\mu$ into (\ref{Bog_met}).

\section{The Friedman equation for a linearized vector field}
The calculation of the curvature and the Ricci tensor leads to the construction of the Friedman equation of motion
for a locally anisotropic universe. We can approximate the 0-component of the primordial-spurionic vector field $n^\mu$ at first order approach
\begin{equation}
n(t)\sim At+B. \label{ntl}
\end{equation}
The special form (\ref{ntl}) is a Taylor type approximation
\begin{equation}
n(t)=n(t_0)+\dot{n}(t_0)(t-t_0)+O( (t-t_0)^2 ). \label{Tayl}
\end{equation}
where
\begin{equation}
A=\dot{n}(t_0),\:\: B=n(t_0)-t_0\dot{n}(t_0).
\end{equation}
Since all of the other components of the spurionic vector field  vanish, only the diagonal elements of the metric and the
Ricci tensor  survive. Under the assumption of a weak Lorentz violation we can restrict our parameter $A$ to be small
enough
\begin{equation}
A=\dot{n}(t_0)\rightarrow 0
\end{equation}
considering an almost constant value of the field.\newline
{\it Connection and curvature}\newline
In virtue of the metric (\ref{osc_metr}), we are able to calculate the Christoffel symbols and the curvature
(see Appendix A).
The Ricci tensors $L_{\mu\nu}$ can be approximated for $b\rightarrow 0, A\rightarrow 0$ and this implies the following
components
\begin{equation}
\begin{array}{rl}
L_{00}= & 3\,\frac { \ddot{a }  }{a }+3\frac{Ab}{B}\frac{\dot{a}}{a} +O \left( {A}^{2} \right)  \\
\: & \: \\
L_{11}= & -\frac{ a\dot{a}+2\dot{a}^2+2k    }{1-kr^2}+\frac{5A}{B}b\frac{a\dot{a} }{1-kr^2}+O(A^2)  \\
\: & \: \\
L_{22}= & -r^2( a\ddot{a}+2\dot{a}^2+2k     )- \frac{5A}{B}br^2a\ddot{a}+O(A^2) \\
\: & \: \\
L_{33}= & -r^2(a\ddot{a}+2\dot{a}^2+2k)\sin^2\theta-\frac{5A}{B}br^2 a\ddot{a}\sin^2\theta +O(A^2)
\end{array}\label{riccit}
\end{equation}
The extra terms appearing at the Ricci components are the dominant ones since they are multiplied by the parameter $Ab/B$
where  $B$ is necessarily small to enable the curvature additional terms to be measurable ($|n^\mu|\ll 1$ ). Following the
usual procedure we can construct the equation of motion for the scale
factor  $a(t)$\cite{Carol}
\begin{equation}
\left(\frac{\dot{a} }{a}\right )^2+\frac{k}{a^2}+2\frac{A}{B}b \frac{ \dot{a} }{a} =
\frac{8\pi G}{3} \left[ \mu -2\frac{A}{B}b P\left( t +\frac{B}{A}\ln B \right) \right] \label{Friedrad}
\end{equation}
The (\ref{Friedrad}) form of the Friedmann equation may not be well suited for late time acceleration since the pressure
term
is a sign of an early universe regime, where the linearization approach possibly breaks down. As the universe evolves we
expect the anisotropic nature of space-time to be converted to a smoother structure. Therefore the linearized approach
of the present model is more convenient for a matter-dominated phase.
\newline
{\it The equation of motion for a matter dominated universe }\newline
Taking into account that there is no  pressure in a matter dominated universe we can obtain the following equation of motion
\begin{equation}
\left(\frac{\dot{a} }{a}\right )^2+\frac{k}{a^2}+2\frac{A}{B}b \frac{ \dot{a} }{a} =\frac{8\pi G\mu}{3}. \label{Fried}
\end{equation}
 We emphasize the extra contribution generated by the geometrodynamical term $2\frac{A}{B}b\frac{ \dot{a} }{a}$.
If the sign of the parameter $Ab/B$ is fixed up to be negative,
the extra term at the equation of motion (\ref{Fried}) will compete dark energy creating a self-accelerating universe .
Despite the fact that the additional term creates acceleration and might replace dark energy contributions, it cannot
 give an answer to the question ``why the vacuum does not gravitate?''. This difficulty gives rise to
the complicated task of distinguishing modifications of curvature from dark energy \cite{Mart_NEB}. However,
the extra accelerating term can be contemplated as a relic left back by an earlier phase where the finslerian
geometrodynamics was characterized by  a non-linear nature ($n(t),b(t)  $  relatively large ). The same Friedman equation
is also produced for the DGP Cosmology for a spatially flat space-time  with
a different continuity equation \cite{defg}. We also remark that the substitution $z_t=2\frac{A}{B}b$ reveals
a correspondence of the present model to the FRW cosmological model described in \cite{rwfins} constructed by a
Randers-Finsler type metric function.
 \newline
{\it Matter density and continuity equation}\newline
The continuity equation for the cosmological fluid of the universe can be directly produced by the conservation law
of the energy-momentum tensor $\nabla_{\nu}T^{\mu\nu}=0$ where the covariant derivative comes from the connection of the
osculating Riemann space-time. The zero component of the conservation law $\nabla_{\nu}T^{0\nu}=0$ implies
\begin{equation}
\begin{array}{l}
\dot{\mu} + \left(2A^0_{00}+\sum_{i=1}^3A^i_{0i}\right)\mu +\dot{P}\left(1-g^{00}\right)+ \\
+P\left[(1-g^{00})\left(2A^0_{00}+\sum_{i=1}^3A^i_{0i}\right)+\sum_{i=1}^3g^{ii}A^{0}_{ii} -\dot{g}^{00}  \right]=0. \label{cont_As}
\end{array}
\end{equation}
 If we make use of the equation of state $w=P/\mu$ and the connection components calculated in Appendix B, the approximation for $A\rightarrow 0$ leads to
\begin{equation}
 \dot{\mu}+3\frac{ \dot{a} }{a}\mu+5\frac{Ab}{B}\mu+\frac{w}{ w+2Ab/B\cdot t+2b\ln B } \Phi [t,a,\dot{a};b,A,B]\mu=0\label{cont_full}
\end{equation}
where $\Phi $ is a function of time, the parameters and the unknowns $a,\dot{a}$. In case of a matter-dominated universe,
where $w=0$, the differential equation can be integrated and give back the solution
\begin{equation}
\mu (t)\propto a^{-3}(t)\exp\left(\frac{-5Ab}{B}t\right)\label{solut_cont}
\end{equation}
which is in alliance to the one found in \cite{rwfins}.
\newline
{\it The Friedman equation in terms of $\Omega$'s}\newline
The whole picture of the cosmological model can be effectively depicted by exploring the
relation of the $\Omega$ parameters.
Using the usual definitions of the Hubble parameter and  the $\Omega$ parameters, we can rewrite (\ref{Fried}) to the form
\begin{eqnarray}
H^2+\frac{k}{a^2}+2\frac{Ab}{B}H=\frac{8\pi G \mu}{3} \label{Hfred}
\end{eqnarray}
which implies the equation
\begin{equation}
\Omega_M+\Omega_K+\Omega_X=1
\end{equation}
where
\begin{equation}
\Omega_X=-2\frac{A}{BH}b
\end{equation}
is the density parameter produced by the extra term of the Friedman equation. The term $\Omega_X$ might give a
significant contribution to the acceleration compared to the dark energy  parameter $\Omega_{\Lambda}\simeq 0.7$.
 \newline
{\it Order of magnitude of unknown parameters}\newline
Since the r.h.s. of (\ref{Hfred}) is positive we restrict $H^2>|2\frac{Ab}{B}H|$.  We can estimate the order of magnitude of the quantity $\frac{Ab}{B}$ in case it dominates the expansion over $\Lambda$
where
\begin{equation}
\frac{Ab}{B}\sim \frac{\Lambda c^2}{6H}.
\end{equation}
Given a typical value of the Hubble parameter
$H_0\simeq  71km/s/Mpc\sim 10^{-18} sec^{-1} $  and the cosmological constant $\Lambda\sim 10^{-57} cm^{-2}$ \cite{Grav}
we deduce
\begin{equation}
\mid\frac{Ab}{B}\mid \sim 10^{-19} sec^{-1} \label{pmest}
\end{equation}
measured in Hubble units.

\section{Energy evolution}
The calculation of the connection components coming from the osculating metric (\ref{osc_metr}) can give us a picture of
how the energy of a particle is affected by the universe expansion and the extra parameters introduced into the
metric function (\ref{BogRW_met_fun}). \newline
{\it Energy of a massless particle }\newline
The 4-momentum of a massless particle is defined by $P^{\mu}=\frac{dx^\mu}{d\lambda}$ where $P^0=E$ is the energy of the
particle. The parameter $\lambda$ is the evolution parameter of the particle's path described by the geodesic equation
\begin{equation}
\frac{d^2x^\mu}{d\lambda^2}+A^{\mu}_{\alpha\beta}\frac{dx^\alpha}{d\lambda}\frac{dx^\beta}{d\lambda}=0. \label{geodn}
\end{equation}
and the zero component of the geodesic equation yields \cite{dod}
\begin{equation}
\begin{array}{rl}
E\frac{dE}{dt}=& -A^0_{\alpha\beta}P^\alpha P^\beta \\
              \: & \: \\
              =& -\frac{Ab}{B}\left(E^2-a_{ij}P^iP^j \right)+(1-b)\frac{ \dot{a} }{a} a_{ij}P^iP^j .
\end{array}\label{energyDEn}
\end{equation}
The particle is massless, $m=0$ thus the dispersion relation (\ref{disp}) is simplified to the usual form
$E^2=-a_{ij}P^iP^j$ hence (\ref{energyDEn}) implies
\begin{equation}
\frac{1}{E}\frac{dE}{dt}-(b-1)\frac{\dot{a}}{a}=-\frac{2Ab}{B}+O(A^2)
\end{equation}
which can be integrated directly and give back
\begin{equation}
E(t)\propto a^{b-1}(t)\exp\left(-\frac{2Ab}{B}t\right).\label{Et}
\end{equation}
The solution (\ref{Et}) possesses an additional redshift effect due to the Lorentz violations inherited by the parameter $b$ and the spurion vector field. The solution behaves  as $E(t)\propto 1/a(t)$ if the extra terms at the equation of motion (\ref{Fried}) are negligible.
\newline
{\it Energy of a massive non-relativistic particle }\newline
We consider a massive non-relativistic particle of mass $m$ traveling on a
geodesic of the space-time
\begin{equation}
\frac{d^2x^\mu}{d\tau^2}+A^{\mu}_{\alpha\beta}\frac{dx^\alpha}{d\tau}\frac{dx^\beta}{d\tau}=0. \label{geod2}
\end{equation}
The 4-momentum of the particle is defined in natural units $(c=1)$ as $P^\mu=m\frac{d x^\mu}{d\tau}$ where $\tau$ is the
particle's proper time.  The chain rule applied to the derivative $\frac{d}{d\tau}$
implies
\begin{equation}
\frac{d}{d\tau}=\frac{E}{m}\cdot\frac{d}{dt}.
\end{equation}
In virtue of (\ref{disp}) the zero component of the (\ref{geod2}) leads to the differential equation for the energy
\begin{equation}
E\cdot\frac{dE}{dt}= -\frac{Ab}{B}E^2
\end{equation}
since $a_{ij}P^iP^j/m^2\rightarrow 0$ for a massive non-relativistic particle . The integration of the differential
equation implies
\begin{equation}
E=E_0\exp\left(-\frac{2Ab}{B}t\right).
\end{equation}
We can calculate the effect of geometrodynamics by measuring the amount of time for a variation of energy 0.1$\:^o/_oE_0$
and find $t\sim 10^{15}sec$ where we used the $\frac{Ab}{B}$-estimation from (\ref{pmest}).

\section{The generalization of the Klein-Gordon equation }
 Since the geometry of  space-time is determined by the process of osculating a Finslerian space to a Riemannian one, the metric is only position dependent. As long as the  velocity is fixed up to be $y\equiv y(x)$, a Riemannian  metric is defined.  Therefore we can apply the General Covariance Principle and construct a curved version of the Klein-Gordon equation (\ref{KGd})
\begin{equation}
 \Box\phi+ m^2(1-b^2) \left(\frac{-in^\mu}{ m(1-b) }\nabla_{\mu} \right)^{2b/(1+b)}\phi=0. \label{curvKG}
\end{equation}
The $ \nabla_\mu$ operator is the covariant derivative coming from the  $A^{\rho}_{\mu\nu}$ connection (see Appendix B), the box operator is
\begin{equation}
\Box\phi=g^{\mu\nu} \nabla_\mu\nabla_\nu\phi=g^{00}\ddot{\phi}-g^{\mu\nu}A^{0}_{\mu\nu}\dot{\phi}
\end{equation}
and $n^\mu$ is the spurion defined in (\ref{nt}) and (\ref{ntl}). We impose the scalar field  $\phi\equiv \phi (t)$ due to the assumption of homogeneity with respect to weak Lorentz
violations. After expanding for small values of $b$ we end up to the following approximation
\begin{equation}
\Box \phi =\left( 1-\frac{2Ab}{B}t \right)\ddot{\phi}+ 3H\dot{\phi}\left(1-\frac{2Ab}{B}t
+\frac{2Ab}{3BH}\right)+O(A^2). \label{box_aprox}
\end{equation}
 A small $A$ and $b$ approximation for the
$m^2(1-b^2) \left(\frac{-in^\mu}{ m(1-b) }\nabla_{\mu} \right)^{2b/(1+b)}$ operator
\begin{equation}
\begin{array}{rl}
m^2(1-b^2) \left(\frac{-in^\mu}{ m(1-b) }\nabla_{\mu} \right)^{2b/(1+b)}\phi=& \: \\
                                                                            =&  m^2\phi+2m^2b\left( \hat{D}\phi \right)+\frac{2bA}{B}t\cdot m^2\phi+2b\ln B\cdot m^2\phi+O(A^2)
\end{array}\label{op_aprox}
\end{equation}
leads to the kinematical equation of the scalar field
\begin{equation}
\ddot{\phi}+3\left(H+\frac{2Ab}{3B}\right)\dot{\phi}+m^2\phi\left(1+\frac{4Ab}{B}t+2b\ln B \right)+O(A^2)=0. \label{KG}
\end{equation}
where $2m^2b \hat{D}\phi\rightarrow 0$ since $b\rightarrow 0$ faster than the logarithmic term
$\hat{D}\phi$. The time derivatives of the
scalar field come from the covariant version of the box operator, therefore the potential
of the scalar field is
\begin{equation}
V(\phi,t)=\frac{1}{2}m^2\phi^2\left(1+2\frac{A}{B}bt+O(A^2)\right).\label{potapp}
\end{equation}
We can eliminate  $H$ if we
combine the Klein-Gordon equation with the Friedman equation of motion producing a differential equation for the scalar field $\phi$.

The energy-momentum tensor is expressed by \cite{ruscosm}
\begin{equation}
T^\alpha_\beta =g^{\alpha \gamma}\phi_{,\gamma}\phi_{,\beta}-\left(\frac{1}{2}g^{\mu\nu}\phi_{,\mu}\phi_{,\nu}-V(\phi,t) \right)\delta^\alpha_\beta
\end{equation}
and determines directly the energy density and the pressure
\begin{equation}
\begin{array}{rl}
\mu =& \frac{1}{2}g^{00}\dot{\phi}^2+V \\
P=& \frac{1}{2}g^{00}\dot{\phi}^2-V. \label{mPphi}
\end{array}
\end{equation}
Thus we can insert (\ref{mPphi}) into (\ref{Friedrad}) and find the Friedman equation of motion
\begin{equation}
H^2+\frac{2Ab}{B}H=\frac{8}{3}\pi G\left[\frac{1}{2}\dot{\phi}^2(1-2b\ln B)+\frac{1}{2}m^2\phi^2\left(1+2b\ln B+4\frac{Ab}{B}t\right)+O(A^2)\right]\label{friedphi}
\end{equation}
The elimination of $H$ from (\ref{KG}), (\ref{friedphi}) yields
\begin{equation}
\begin{array}{l}
\ddot{\phi}+\sqrt{12\pi G}\left(\dot{\phi}^2+m^2\phi^2\right)^{1/2}\dot{\phi}+m^2\phi+ \\
+ \frac{Ab}{B}\cdot  f(t,\phi,\dot{\phi})+2b\ln B \cdot g(t,\phi,\dot{\phi})+O(A^2) =0
\end{array}
\label{phiode}
\end{equation}
where
\begin{equation}
\begin{array}{rl}
f(t,\phi,\dot{\phi})=& -\dot{\phi}+4m^2t\phi + \\
                    +& 2m^2t\phi^2\dot{\phi}\sqrt{12\pi G}\left(\dot{\phi}^2+m^2\phi^2\right)^{-1/2}
\end{array}\label{fKGfrd}
\end{equation}
and
\begin{equation}
g(t,\phi,\dot{\phi})=m^2\phi-\frac{1}{2}\sqrt{12\pi G}\left(\dot{\phi}^2+m^2\phi^2\right )^{-1/2}\left( \dot{\phi}^2-m^2\phi^2 \right ) \dot{\phi}
\end{equation}
 are functions of $\phi$ and its first derivative ( for details see e.g. \cite{ruscosm} ) .
\newline
{ \it The limit $ |\dot{\phi}| \ll  m\phi $ }\newline
In this special case the potential is much larger compared to the kinetic energy thus we can derive the
simplified differential equation

\begin{eqnarray}
\ddot{\phi}+m^2\left(1+4\frac{Ab}{B}t+2b\ln B \right)\phi=0. \label{phiL}
\end{eqnarray}
The solution of (\ref{phiL}) is expressed with the aid of Airy functions \cite{AbStg}
\begin{equation}
\phi(t)=C_1Ai\left(\zeta \right)+C_2Bi\left( \zeta\right)\label{Aisl}
\end{equation}
where
\begin{equation}
\zeta= \left( -2m\frac{B}{Ab} \right)^{2/3}\left(-\frac{1}{4}-\frac{Ab}{B}t-\frac{b}{2}\ln B \right).
\end{equation}
 The argument of the Airy function is fixed up $\zeta\sim -\frac{1}{4} \left( -2m\frac{B}{Ab} \right)^{2/3}$, given the
small values of the parameter $\frac{Ab}{B}$, hence we can regard $\zeta$ to be independent of time. We can approximate the
function $\phi(t)$ using the asymptotics of the Airy function after taking into
consideration the negative sign of argument $\zeta$ \cite{AbStg} .  The substitution of (\ref{Aisl}) to the
Friedman equation (\ref{friedphi}) yields
\begin{equation}
a(t)= \exp \left ( -\frac{Ab}{B} t \right ) \exp [ h(t) ] \label{infl}
\end{equation}
where
\begin{equation}
h(t)=\frac{2}{3}C_3(\zeta)^{1/2}\left( 1+\frac{Ab}{B}t+2b\ln B \right)^{3/2}
\end{equation}
Given the positive sign of $\ddot{a}(t)$ in (\ref{infl}), the solution secures an inflationary phase for a time interval
where the high potential takes over the expansion.

\section{Discussion}
The essential result of our approach is the adoption of a Finslerian metric function
 applied to Cosmology, coming from Cohen and Glashow's very special relativity \cite{Coh}. The calculation of the
Einstein's field equations for an osculating Riemannian space-time gives back a Friedman equation for a self accelerating
universe under the assumption  that the sign of the extra parameter is negative. The
estimation of the energy evolution implied by the modified geodesic equations may lead to experimental constraints for the
VSR theory using observational data from the large scale structure. The specific limit for a massive non-relativistic object
implies a small variation of energy within a period of time close to the age of the universe restricting our calculations
within the acceptable observational limits even in case the model wins totally over the dark energy.

The construction of the kinematical equation of a scalar field \cite{Gibb_fins} with the aid of the Finslerian metric
(\ref{osc_metr})
and the Friedman equation (\ref{Fried}) can lead to a better understanding of the nature of the Finslerian
 gravitational field. Indeed, the Lorentz violations provide a modified potential for the curved Klein-Gordon
equation (\ref{KG}) affecting the validity of the strong energy conditions.  Given  a large potential compared to the
kinetic energy, the solution of the scale
factor implies an inflating phase of the expansion depended on the GVSR assumptions inherited by the metric function
(\ref{BogRW_met_fun}) . An interesting task for future work is the study of the model's early time behavior producing
 an inflationary solution without the aid of a scalar field, considering stronger effects of Lorentz symmetry
breaking. A similar  inflationary scenario has also be produced
by gravitational mechanisms, as a direct result of Lorentz violations not depending on the vacuum's fluctuations
and GUTs \cite{gasp}.

An explanation of acceleration lies on the fact that dark energy acts as a repulsive force introducing a
 cosmological constant at the Einstein field equations. In such a case the cosmological constant
is a finely tuned ground state of a potential implying negative pressure at the equations of motion . A universe with
pressure free matter can be self accelerating under
the restriction of a modified Ricci curvature which imposes an asymptotically de Sitter geometry of space-time.
The  machinery of osculating a Finslerian space to a Riemannian one, leads at first order approach directly to
an asymptotically de-Sitter universe. However, the classification of the present model as a low curvature modification (e.g.$\Lambda$CDM, DGP) needs to be proven. This is a vital task since all such cosmological models reproduce Newtonian
gravity locally \cite{devsdg}.

A possible estimation of the spurionic vector field (from high energy physics or other methods), within the limits
of the present cosmological model, can set forth an answer to the vital question about the small value of $b$
posed by Gibbons et.al (see \cite{Gibb_fins} ), connecting Lorentz violations to {\it dark energy} problem. A further
investigation of the present model taking into account the calculation of cosmological perturbations and the CMB data
may relate Lorentz violations to the problem of large angle anisotropies and inhomogeneities.

The introduction of Finsler geometry as a geometry of space-time opens up a new direction towards the study of geometric
phase transitions. The concept of geometric phase transitions generated by the Bogoslovsky's line element (\ref{Bog_met})
has already been studied in \cite{Bogoslovsky} for the special case of a flat Finslerian space-time. An interesting
generalization can be applied to a curved Finsler space for a better understanding of how Lorentz violations, with a
varying $b\equiv b(t)$, may evolve as the universe expands. Since Lorentz violations produce anisotropies, it is
natural for them to ``dilute'' to thermal energy and large amount of entropy \cite{Grav}, therefore the special limit
of the present model will be asymptotically recovered.

\section{Acknowledgments}
We wish to thank the University of Athens (Special Accounts for Research Grants) for the support to this work and M.Gasperini for his encouraging comments.

\appendix

\section{The metric $g_{\mu\nu}$}
The definition of the Finsler metric function (\ref{BogRW_met_fun}) implies the calculation of the space-time's metric.
Assuming the comoving character of the spatial coordinates and the vector field tangent to the cosmic flow lines the
computation of the metric components of the osculating Riemann space-time are simplified to the form ($i,j=1,2,3$ and $a_{\mu\nu}$
 is the Robertson-Walker metric (\ref{RW}) )
\begin{equation}
\begin{array}{rl}
g_{00}=& 1+2\ln (At+B)b+O(b^2) \\
g_{ij}=& a_{ij}+a_{ij}\left[2\ln (At+B)-1\right]b+O(b^2)
\end{array}\label{metric_bapp}
\end{equation}
which implies the connection $A^{\rho}_{\mu\nu}=g^{\rho\sigma}\frac{1}{2}\left(g_{\sigma\mu,\nu}+g_{\sigma\nu,\mu}
-g_{\mu\nu,\sigma} \right)$ and the curvature $L^{\mu}_{\nu\alpha\beta}=A^{\mu}_{\nu\beta ,\alpha}-A^{\mu}_{\nu\alpha ,\beta}
+ A^{\mu}_{\sigma\alpha}A^{\sigma}_{\nu\beta} - A^{\mu}_{\sigma\beta}A^{\sigma}_{\nu\alpha}   $. The Ricci tensor
components (\ref{riccit})  are calculated for the limit $A\rightarrow 0$.

\section{ The connection components $A^{\rho}_{\mu\nu}$ }
\begin{equation}
\begin{array}{rl}
A^{0}_{00}=& \frac{Ab}{B}+O(A^2)  \\
\: & \: \\
A^{1}_{01}=& A^{2}_{02}=A^{3}_{03}= \frac{\dot{a} }{a}+ \frac{Ab}{B}+O(A^2) \: \\
\: & \: \\
A^{0}_{ij}=&-(1-b)\frac{\dot{a}}{a}a_{ij}-\frac{Ab}{B}a_{ij}+O(A^2) \\
A^{1}_{11}=& \frac{kr}{1-kr^2}  \\
\: & \: \\
A^{1}_{22}=& -r(1-kr^2)  \\
\: & \: \\
A^{1}_{33}=& -r\sin^2\theta(1-kr^2)  \\
\: & \: \\
A^{2}_{12}=& A^{3}_{13}=\frac{1}{r} \\
\: & \: \\
A^{2}_{33}=&-\sin\theta\cos\theta \\
\: & \: \\
A^{3}_{23}=& \cot\theta
\end{array}
\end{equation}

\end{document}